\documentclass{article}
\begin{document}

\title{Reductionism, Emergence,\\ and Effective Field Theories}
\author{Elena Castellani\thanks{Department of Philosophy, University of 
Florence, via Bolognese 52, 50139 Firenze, Italy. E-mail: castella@unifi.it}}
\date{April 2000}
\maketitle

\begin{abstract}
In recent years, a ``change in attitude'' in particle physics has led to our 
understanding current quantum field theories as {\it effective field theories}
(EFTs). The present paper is concerned with the significance of this EFT approach, 
especially from the viewpoint of the debate on reductionism 
in science. In particular, it is  
a purpose of this paper to clarify how 
EFTs may provide an interesting case-study in current philosophical
discussion on reduction, emergence and inter-level relationships in general.
\end{abstract}

\bigskip


\section{Introduction: A Recent ``Change in Attitude'' in Particle Physics}

According to a view dominant in the 1970s and 1980s 
and still widely accepted, ``fundamental physics'' is the physics concerned 
with the search for the ultimate constituents of the universe and the laws 
governing their behaviour and interactions. Fundamentality, on this view, is  
the prerogative of the physics of smaller and smaller distances 
(or higher and higher energies), and accordingly particle physics and 
cosmology are currently identified as the fields where the quest for the 
``final theory'' takes place. 

Historically, the rise and establishment in the 1970s of the Standard Model 
(electroweak theory and quantum chromodynamics)  marked a pivotal step in this 
search: a description of the physical world was obtained 
in terms of quarks and leptons as basic constituents of matter and 
their (nongravitational) interactions, interpreted in a unified way as 
controlled by the same principle (the gauge principle) and all described 
by a renormalizable quantum field theory. Among the many implications 
connected with the success of the Standard Model (tested 
down to distances of 10$^{-18}$ cm, in some cases), let us stress here 
two points. First, because of the apparently unified and fundamental character 
of the physical description it provided  --- 
a few basic entities and a unified treatment of their interactions ---
the Standard Model was commonly seen as 
representing a major success of reductionism and unification in 
natural science (Schweber, 1993, p. 35). Second, although the 
Standard Model was clearly not the whole story (since it left out gravity and 
also relied on a number of arbitrary parameters), its theoretical success 
and many empirical confirmations made it quite natural to expect  
fundamental physical theories to have the shape of a 
renormalizable quantum field theory (Weinberg, 1999, p. 242).

In more recent years, further developments in quantum field theory  
and in the application of renormalization group theory have brought  
a change in this way of regarding current fundamental physical theories.
As a result of a ``change in attitude'' in particle physics 
(Weinberg, 1997, p. 41),  
the Standard Model is now understood as an {\it effective field theory} (EFT), 
that is, the low-energy limit of a deeper underlying theory which may 
not even be a field theory. Once one views current quantum field 
theories as effective theories, different positions about 
what is {\it physically fundamental} are possible. In recent years, 
the following two positions have emerged:

\begin{enumerate}

\item[$\bullet$] The view that there is a fundamental theory --- 
the so-called theory of everything (superstring theory, M-theory, ...) --- 
at the bottom of both the Standard Model and general relativity (each understood 
as effective theories). This is the view prevailing among high energy physicists 
(S. Weinberg, E. Witten, D. Gross,...) and commonly labelled as ``reductionist''. 

\item[$\bullet$] The view that physics obtains at most 
a never-ending tower of EFTs, with the corresponding picture of the physical
world as layered into quasi-autonomous domains, each level having its own 
ontology and its own fundamental laws. The view, labelled as 
``antireductionist'', was formulated in such terms and especially defended  
by S. S. Schweber and T. Y. Cao in a series of recent works (for example: 
Cao and Schweber, 1993; Schweber, 1993; Cao, 1997, and 1999).  

\end{enumerate} 

As regards fundamentality, two main questions are at stake here, 
namely: (a) whether there is a fundamental theory  
of everything (the ``final theory''), and (b) what it means to say 
that a theory is ``more fundamental'' than another.
Discussions of such issues usually take place in the context
of the {\it debate on reductionism in science}; where the questions (a) and 
(b) are commonly addressed in connection with, respectively, 
(a') the question of whether everything is reducible to a 
fundamental level (theory), and (b') the question of the nature 
of inter-level relationships (inter-theory relationships) in a hierarchy
of levels (theories).

The various implications of
the EFT approach in particle physics, its philosophical significance and in 
particular its relevance to such themes as reductionism and 
fundamentality in science are now at the center 
of a developing debate among physicists, historians and philosophers of 
physics (see for example Cao, 1999). A first aim of this paper is to clarify the 
terms of this debate, by exploring its 
background, the main concepts and issues involved and the key controversial 
points. The concern, here, is not so much with the details of the different 
personal positions, as with the general ideas that come into play. 
Accordingly, attention is paid more to the conceptual than  
to the ideological aspects (however effectively important ideologies are).

The first part of the paper is concerned with the historical background to 
the debate on EFTs, namely the discussion 
on reductionism and fundamentality in natural science as it developed 
in the scientific community 
from the middle 1960s to the early 1990s. As shown in Section 2, the 
majority of the basic issues addressed in relation 
to EFTs emerge in the context of this discussion: 
the present EFT debate can actually 
be seen as the most recent phase of such a discussion,  
its natural evolution in the light of new scientific developments and 
physicists' ``changes in attitude''.
In the conceptual background to the whole subject there is obviously also
the general philosophical debate on such notions as reduction, 
supervenience and emergence, and on related doctrines like reductionism,
emergentism, and physicalism. 

But curiously enough, despite the 
growing need in this context to pay more attention to what happens in  
basic science and especially in today's physics 
(a point considered in Section 3), the literature does not seem 
to contain specific studies of the connections between the strictly philosophical 
debate and the  ``scientists' debate'' of the last years. 
It is therefore one main purpose of this paper to show that EFTs do 
provide a new and surely interesting case-study for the current philosophical 
discussion on reduction, emergence and 
inter-level relationships in general. The last two Sections are devoted 
to this task. Starting with the general idea of an EFT and its actual 
significance in particle physics (Section 4), some main controversial points 
of the debate on the EFT approach are considered in the final Section, where 
a conclusion is attempted as regards what EFTs can suggest and what they 
actually do not imply.

\section{Funding  
and the Nature of Basic Science: A Debate among Scientists}

In the early 1960s, for a number of reasons that are clearly 
illustrated in a recent article by Silvan Schweber (1997), 
the question of how to allocate finite government 
research funds started to become an urgent one in the United States. 
Accordingly, what criteria ought to be followed in making the funding 
choices among the different fields of science became the subject of a 
growing debate involving science administrators, economists, scientists 
and also philosophers. The problem was particularly felt
with regard to basic research and especially high energy physics (HEP): 
why spend such a large amount of money, as required to support programs 
of experimental particle physics using accelerators, on a field far removed 
from our daily experience and therefore much less likely to be of immediate 
technical and practical relevance than other branches of science? 
Funding, in this case, could not be 
justified on the grounds of important contributions to everyday 
technology and human welfare.
Reasons for spending money on HEP had to be found ``internally'', 
by looking at the nature, purposes and value 
of particle physics.\footnote{Following the common usage, I shall use 
``HEP'' and ``particle physics'' indifferently here.}
Thus a discussion arose, especially among 
practicing scientists (mainly physicists); which can be taken as  
the starting point of the ``scientists' debate'' on reductionism 
and fundamentality in natural science representing, as we shall see, 
the historical background to the present debate on EFT.

Let us make a selection and, following a rough
periodization, single out two 
phases of this scientists' debate that are of special relevance from 
the viewpoint of our paper, as follows. 1) A first phase,    
ranging from the mid-1960s to the rise of the Standard Model in 
the early 1970s, which was centered on the topic of 
large public support for the building of new particle 
accelerators and which began with the challenge posed by the physicist Alvin 
Weinberg (at the time director of the Oak Ridge National Laboratory) 
to the value of HEP and to the criteria 
for choice of which scientific research programs to pursue in general 
(A. Weinberg, 1963 and 1964). 
2) A second phase, following the period of enthusiasm for the 
success of the Standard Model, which was dominated by the discussion 
of the Superconducting Super Collider (SSC) --- a scientific 
project that started to be planned in 1982 and was definitively cancelled 
at the end of 1993.

1) Conceptually, a central topic in the 1963-1964 papers 
by Alvin Weinberg and the 
more or less direct responses to the problems they posed was the 
following question: what is the intrinsic value 
of basic research such as particle physics?
The most common argument in support to HEP proposed at the time 
(but often also today) is the one identifying the intrinsic value of this field 
with its ``fundamentality''. Particle physics is more fundamental 
than other areas of science (physics) in the sense that it contributes most 
to the advance of our fundamental understanding of the physical world, it is on 
the frontier of the exploration of nature to all its limits, it moves us 
closer to the ultimate laws of nature or `to the absolute logical structure 
of the universe' (S. Weinberg, 1965, p. 73): this is the sort of 
statement to be found in the 1965 volume 
edited by Luke C. L. Yuan (at the time, the chairman of the 
Committee for the building of a new accelerator at Brookhaven),  
containing the views on the objectives of particle physics 
of a number of physicists such as Hans A. Bethe,  
Gary Feinberg, Julian Schwinger, Victor F. Weisskopf and Steven Weinberg. 
``Fundamentality'', ``frontier research'' and ``unification'' 
are the key concepts recurring in this line of defense of 
HEP.\footnote{The unification issue and its relevance to the scientists' debate 
on reductionism are thoroughly examined in Cat (1999).} 
A further articulation was provided by Weisskopf's distinction between two 
trends in the development of twentieth century science, namely: 
 `intensive research', including HEP and a 
good part of nuclear physics, that `goes for the fundamental laws'; and 
 `extensive research', such as solid state physics and plasma physics,  
that `goes for the explanation of phenomena in terms of known fundamental 
laws' (Weisskopf, 1965, p. 24).  

The fundamentality argument in support of HEP as well as 
the intensive/extensive distinction were especially challenged by 
the leading solid state physicist 
Philip W. Anderson in a seminal article with the programmatic title 
{\it More Is Different} (1972). 
Anderson's article, now considered a ``classic'', was expressly written to 
defend the intrinsic value of other branches of physics, such as 
solid state physics (today: ``condensed matter physics'') and in general what 
goes under the name of ``low energy physics'' (LEP),  
against a presumed intellectual (and therefore, financial) 
supremacy of HEP. Its main purpose was to oppose 
`what appears at first sight to be an obvious corollary of reductionism: 
that if everything obeys the same fundamental laws [the reductionist 
hypothesis], then the only scientists who are studying anything really 
fundamental are those who are working on those laws' (p. 393).
Anderson's task was therefore to bring arguments to 
demonstrate how the work, say, of a solid state physicist could be 
as fundamental as the work of a particle physicist. In view of the 
relevance of his paper to the ``scientists' debate'', 
especially with regard to its successive developments,
let us recall some basic notions on which his 
arguments rely:\footnote{How 
philosophically sound was Anderson's way of arguing in his 1972 paper 
will not be our point here. 
For such a sort of analysis see for example Cat (1999, pp. 264-266).}  

a) {\it The way down/way up distinction}.\footnote{I am using 
here the terminology typically employed in the more recent 
phase of the scientists' debate.} Although commonly considered as 
an outstanding example of an antireductionist, Anderson does not oppose 
in his paper what he calls the {\it reductionist hypothesis}, `the ability to 
reduce everything to simple fundamental laws' (p. 393) --- `we must all 
start with reductionism, which I fully accept' (p. 394) --- but rather the 
{\it constructionist hypothesis}, that is `the ability to start from 
those [fundamental] laws and reconstruct the universe' (p. 393).

b) {\it The fact of emergence}. The constructionist hypothesis is rejected 
because of the emergence of new properties: 
`at each new level of complexity entirely new properties appear, 
and the understanding of the new behaviors requires research 
which I think is as fundamental in its nature as any other' 
(p. 393).\footnote{Anderson's paper especially 
stresses the relevance of broken symmetry to the fact of emergence. 
This important point is considered in another paper (Castellani, 
`Spontaneous Symmetry Breaking and Emergence', in preparation).}

c) {\it The hierarchical structure of science}. Because of the 
fact of emergence, `one may array the sciences ... in a hierarchy'
according to the idea that the elementary entities of 
(less primitive) science X obey the laws of (more primitive) 
science Y, but this does not imply that science X 
is `just applied Y' (p. 393). Condensed matter physics is not  
applied elementary particle physics, chemistry is not applied condensed 
matter physics, and so on. The entities of X are emergent in the sense that,  
although obedient to the laws of the more primitive level Y 
(according to the reductionist hypothesis), 
they are not conceptually consequent 
from that level (contrary to the constructionist 
hypothesis).\footnote{See the following statement by 
Anderson (1995, p. 2020): `This, then, is the fundamental philosophical 
insight of 
twentieth century science: everything we observe emerges from a more 
primitive substrate, in the precise meaning of the term ``emergent'', 
which is to say obedient to the laws of the more primitive level, but 
not conceptually consequent from that level'.}

d) {\it Decoupling between HEP and LEP}. What happens at the HEP level 
is not very relevant to what effectively happens at the LEP level: 
`the more the elementary particle physicists tell us about the nature 
of the fundamental laws, the less relevance they seem to have to the very 
real problems of the rest of science' (Anderson, 1972, p. 393).\\

2) During the 1980s, the debate on funding and criteria for scientific 
choice took a new turn by focusing --- as regards physics --- on 
a `scientific project of 
unprecedented size and cost' (Weinberg, 1993, p. 1): the building of a 
new particle accelerator known as the Superconducting Super Collider (SSC), 
with the objective of attaining the 1-TeV mass scale, that is,  an energy 
domain where some answers could be found to questions left open by the 
Standard Model (first of all, the mechanism of electroweak 
symmetry breaking).\footnote{To reach 
the required energy for the creation 
of a Higgs particle was surely a central motivation 
for the SSC, but not the only one. This point is especially stressed in 
Weinberg (1993, pp. 210-215). Greater details on the various purposes of 
the SSC project can be found in Glashow and Lederman (1985).} 
Because of the cost of the project 
(estimated at about 4.4 billion dollars in 1987), 
well known scientists were asked to testify at congressional 
committee hearings on the significance of the SSC and therefore ``forced'' 
to take a definite position. Solid state physicists like  
James Krumhansl and Anderson opposed the project, for reasons similar to 
the argument defended in Anderson's 1972 paper, 
while the majority of particle physicists strongly supported it. 
It is not our aim to enter into the details of the SSC debate here. 
Let us just mention --- as best illustrations of a ``reductionist pro-SSC'' 
position, on the one side, and of an ``antireductionist anti-SSC'' 
position, on the other side --- the arguments advanced by  
Steven Weinberg and the 
prominent biologist Ernst Mayr, respectively, in an exchange 
between the two scientists which appeared in the journal {\it Nature} 
in the years 1987-1988.

Weinberg's 1987 defense of the SSC --- and in general of spending on 
``big science'' --- is grounded on the 
usual (in the HEP supporters' camp) argument that 
`particle physics is in {\it some} sense more fundamental 
than other areas of physics' (1987, p. 434). 
In what way? Weinberg's position relies on what he calls {\it objective 
reductionism}: a reductionism described in terms of `the convergence 
of arrows of scientific explanation', `objective' because it is  
`not a fact about scientific programmes, but a fact about nature' (p. 436). 
Behind this view is the `intuitive idea that different scientific 
generalizations explain others [generalizations]': so we 
have `a sense of direction in science', `there are arrows of 
scientific explanation that ... seem to converge to a common 
source' (p. 435). Particle physics, dealing with nature `on a level closer 
to the source of the arrows of explanation than other areas of physics', 
is therefore more fundamental (p. 437).

As noted in the literature, 
a main controversial point in Weinberg's argument is his intuitive idea 
of explanation --- an explanation which can be also only ``in principle'' and 
is generally distinguished from deduction (in order to admit the fact 
of emergence).\footnote{In subsequent works (1993, 1995a, 1998) Weinberg,
challenged on this point, tries to be more precise about what 
he means by explanation and objective reductionism, later dubbed by him 
``grand reductionism'' 
(`the view that scientific principles can all be in principle traced 
down to a small body of simple universal laws' [1998, p. 79])
and opposed to what he calls ``petty reductionism'' (`the view that 
the mere knowledge of the constituents of a complex system would be 
sufficient to understand the system' [ibid.]). We prefer not to enter here 
into the question as to whether Weinberg actually succeeded in making his 
position philosophically sound. In any case, he is surely very clear on 
what he thinks about reductionism and fundamentality 
in natural science.} Weinberg's position actually 
does not easily fit into traditional philosophical schemes, and even  
Mayr, in his 1988 article on the limits of reductionism, encounters some 
difficulties in classifying it from the viewpoint of  
his own 1982 distinction among three kinds of reductionism: namely, 
a {\it constitutive reductionism}, which is just the
analytic method of studying an object by reducing it into its most 
basic consituents; a {\it theory reductionism}, postulating 
that `theories and laws formulated in one field of science can be shown 
to be special cases of theories and laws formulated in some other branch of 
science' (1982, p. 62); and an {\it explanatory reductionism}, 
which is `the view that the mere knowledge of its ultimate components 
would be sufficient to explain a complex system' (1988, p. 475).
According to Mayr, Weinberg is defending something similar to  
theory reductionism; while Weinberg, on his part, prefers to stress 
the objective (not semantic) character of his own view. 

The really ``bad'' reductionism, for Mayr, is explanatory reductionism. 
Mayr defines his antireductionist position essentially in opposition 
to this kind of reductionism, which he rejects on the basis 
of the fact of emergence (just as in Anderson's 1972 paper the 
constructionist hypothesis was rejected on the grounds of emergence): 
`new and previously unpredictable characters emerge at higher levels 
of complexity ... hence complex systems must be studied 
at every level, because each level has properties not shown at 
lower levels' (1982, p. 64). For this reason Mayr is then led to 
doubt that HEP research (the SSC project, in the specific case)
`would contribute not only to our understanding of the 
subatomic world, but also to that of the middle world' (1988, p. 475).

As we see, the fact of emergence, a kind of autonomy of levels of 
organization and hence a sort of ``decoupling'' are central themes in 
the antireductionist camp of the ``scientists' debate''. But also 
a reductionist like Weinberg accepts the fact of emergence and agrees 
in rejecting what Mayr calls explanatory reductionism. So on what does the 
disagreement really rest in this debate? From what we have seen so far, 
on the consequences (we could say: ideological consequences) 
that are drawn from the same facts --- for example, the fact of 
emergence: i.e. the consequences about what is more fundamental in science 
and what is the scope of basic research. This is exactly the same 
situation that we shall find when considering the controversy over the 
philosophical implications of EFT. 
As we shall see in Section 5, from the same facts --- the  
EFT approach in particle physics and its ``extreme version'' ---  
quite different conclusions have been drawn as regards reductionism 
and fundamentality in physics.

\section{Levels and their Relationships: Some Desiderata in 
the Philosophical Debate}

Mayr's three types of reductionism resemble 
categories typically employed in the philosophical debate 
on reduction and reductionism, 
where distinctions between some kinds of reduction (reductionism) 
--- ontological reduction, epistemological or semantic reduction,
 explanatory reduction or 
reductive explanation, methodological reduction, ... --- are ordinarily 
discussed.\footnote{Mayr's classification has also inspired philosophical 
reflection; a good example is the classification of models of reduction 
(`theory reduction', `explanatory 
reduction', and `constitutive reduction') offered in Sarkar (1992), on the 
grounds of distinctions partly based on Mayr's categories.} 
Most concepts recurring in the scientists' debate, such as emergence, 
hierarchy of levels (levels of organization/levels of description), 
analysis {\it versus} synthesis and the 
part/whole relation are in fact current philosophical subject 
matters. But the field of application 
is markedly different: much of the recent philosophical 
discussion on such matters is mainly concerned with the relation of mental 
to physical phenomena. Among philosophers, however, 
it is now becoming a desideratum `to stop thinking 
of these issues exclusively in terms of mental properties, and to look 
for examples in more basic science', as has been said 
with regard to the question of emergent properties (Humphreys, 1997, p. 15). 

Both the scientists' debate of Section 2 and 
the general philosophical debate on reduction, 
supervenience and emergence are in some sense centered on the following 
question: assuming a level structure of ``units'' of some kind, 
how are the levels related? The question, of course, first of all depends 
on what level structure is assumed. Some philosophers prefer not to speak of 
levels altogether; Sahotra Sarkar (1998), for example, suggests using the 
term ``realm'' instead of the more common ``level'', in order to avoid 
assuming a hierarchical relationship between different levels (fn. 22, 
p. 193). Given that we are not so much concerned, in this paper, with 
``intra-level'' or ``domain-preserving'' relationships 
as rather with ``inter-level'' relationships (to use current 
terminology), let us speak freely of levels. Moreover, since the whole 
debate we are considering 
is basically referred to the physical world and its description, 
let us follow the common usage among natural scientists 
and understand ``levels'' in terms of {\it a given scale} ---  
a length scale, for example, or an energy scale. 
A level (and its hierarchical position) will accordingly be intended as 
defined in correspondence to a given value range of the scale: 
in terms of energy, for example, a level  
$L_{i}$ will be considered ``coarser'' than 
another level $L_{j}$ if corresponding to a lower range of 
energy values.

Given a hierarchical structure of two or more levels $L_{i}$'s, 
where level $L_{i+1}$ is coarser than level $L_{i}$, the main 
question is then: how are the units in two successive levels $L_{i}$ 
and $L_{i+1}$ related? Is the relation of the units in level $L_{i+1}$ to 
the units in level $L_{i}$ better described in terms of ``reduction to'', or 
``supervenience on'', or ``emergence from''? Again, some preliminaries 
need to be settled; first of all, what are the appropriate units 
in the level structure: theories, concepts, entities, properties, ...? 
As has been pointed out, much of the philosophical discussion 
has been ``clouded'' by a lack of agreement on this point 
(Klee, 1984, p. 45).\footnote{To be precise, Klee's 
observation is especially referred to the philosophical discussion 
on emergence.}
A way of restricting the unit choice is to start with distinctions such 
as those between ontological and epistemological aspects, 
and formal and nonformal (substantial) issues.\footnote{We do not need to 
enter into such details here. A sort of overview, although not complete,  
of these distinctions in the philosophical literature is offered in 
Sarkar (1998, Chapters 2 and 3).}
Let us cut the discussion short, here, and just take one example:
namely, the recent analysis offered by Jeremy Butterfield 
for reduction, supervenience and emergence as (formal) relations 
holding between {\it theories} (Butterfield and Isham, 1999, 
pp. 114-126).

Intending theories as sets of sentences closed under deduction 
(the syntactic conception),\footnote{According to Butterfield, this choice 
is not decisive: the arguments he presents would hold also under other choices, 
and in particular on the semantic 
conception of theories as classes of models.} 
Butterfield's analysis 
first aims to show that, by approaching reduction 
in terms of ``definitional extension'' (the notion 
making precise the intuitive idea of one theory $T_{1}$ being reduced 
to another theory $T_{2}$), `there may well be no 
single ``best'' concept of reduction' (p. 120). On the one side, definitional 
extension is sometimes too weak for reduction, whence the controversy 
over how to supplement it (starting with Ernst Nagel's well known proposal, 
motivated by the idea that $T_{2}$ should explain $T_{1}$, where explanation 
is conceived in deductive-nomological terms).
On the other side, definitional extension is sometimes too strong (an 
objection going back to Paul Feyerabend), prompting one to appeal to such 
concepts as approximation, limiting relations and analogies between 
theories. Supervenience, 
a notion `apparently weaker than definitional extension but 
also quite precise' and therefore by many considered a good candidate for 
capturing the idea of emergence, is then shown not to live up to its 
promise of sharply distinguishing reduction and 
emergence. Turning finally to emergence itself, what 
results is a heterogeneous picture of this notion.  
Butterfield's conclusion is then the following: rather 
than seeking for a general formal definition of emergence (for which there 
does not seem to be much prospect, as in the case of reduction), 
`we need to bear in mind the variety of ways that theories can be related: 
in particular, with one theory being in some sense a limit of the other, or 
an approximation to it' (p. 115). 

The preceding analysis is a very good example of how the need 
to search for concrete examples and to be `concerned 
with detailed mechanisms rather than broad generalizations' 
(Sarkar, 1998, p. 17) is becoming commonly felt in the 
philosophical discussion on reduction, emergence, and inter-level 
relationships in general. 
It is exactly in this spirit that we shall turn now to consider 
the EFT approach and its significance from the viewpoint of the above 
discussion. A main aim of this paper is to show that EFTs 
do provide a new and stimulating concrete case-study for this discussion.

\section{EFT: The General Idea and its ``Extreme Version''}

As a way of introducing EFT, let us start with the general 
idea of an ``effective theory'' and then see 
how this idea applies in the framework of 
quantum field theory (QFT).\footnote{Historically, it is rather the other 
way round. According to Weinberg (1995b, p. 523), an earliest example 
of an EFT in physics is the theory of low-energy photon-photon interactions 
derived in the 1930s by H. Euler and others. Effective Lagrangians for soft pions 
were used since 1967, but it is actually only in the late 1970s (on the 
grounds, first of all, of important work by Weinberg in 1979) that 
it was realized that `effective field theories could be regarded as 
full-fledged dynamical theories, useful beyond the tree approximation' 
(Weinberg, 1997, p. 42).}

An effective theory (ET) is a theory which ``effectively'' 
captures what is physically relevant in a given domain.\footnote{For the 
scope of our discussion, it will 
be sufficient to employ the term ``theory'' in the usual 
physicists' sense of a set of fundamental equations (or simply some Lagrangian) 
for describing some entities, their behaviour and interactions.} More precisely, 
following a characterization by the particle physicist Howard Georgi,  
an ET is an {\it appropriate} description of the {\it important} (relevant) 
physics in a given region of the parameter space of the physical world 
(1997, p. 88). 

The idea is a very natural one: physics usually changes as one 
changes the scale considered (what is physically relevant differs 
from one region of the parameter space to another);\footnote{As we shall 
see, on the 
renormalization group approach the effect of 
changing the scale can be absorbed in a change of the parameters of the theory.} 
at very different ranges of energy scales (or length scales), 
we can have remarkably different physics.\footnote{In particle physics, as 
is known, energy and length scales are inversely related. The usual 
unit system is $\hbar=c=1$, whence the dimensional relations 
$[length]=[time]=[energy]^{-1}=[mass]^{-1}$. The mass $M$ of a 
particle is therefore equal to its rest energy $Mc^{2}$ and to its 
inverse Compton wavelength $Mc/\hbar$.} 
Which is to 
say, when referring to one physical system, that the system in ``coarse 
grain'' can appear very different from that in ``fine grain''. 
Now, in general, it is possible to describe a limited range of physical 
phenomena, or equivalently the relevant physics in a limited region of 
the parameter space, without having to describe everything at once. 
This is the point of using 
an ET: to obtain a most appropriate and convenient description of the 
relevant physics in a limited domain. 
Such a description, only applicable within a well defined domain of 
validity, is thus intrinsically {\it approximate}.  

For reasons we shall briefly see next, 
this general ET idea proves particularly useful 
and interesting in the framework of QFT, where it can be made 
very precise. Actually, the development of the EFT idea 
has been of special consequence to particle physics, 
leading to the ``change of attitude'' mentioned in Section 1 which 
is at the basis of the so-called modern view of QFT: 
i.e., the view that `the most appropriate description of particle 
interactions in the language of QFT depends on the energy 
at which the interactions are studied' (Georgi, 1989, p. 446). 
On this view, current QFTs are understood as EFTs, 
each EFT explicitly referring only to those 
particles (fields)\footnote{Let us follow theoretical 
physicists' common usage in referring to `particles' as the 
objects of their theories. The controversy about     
what the basic QFT entities really are is not of special relevance 
to the matters being discussed here.} that are actually of importance at 
the range of energies considered. By changing the energy scale 
the EFT description accordingly changes, `to reflect the changes in the 
relative importance of different particles and forces' (ibid.). 

As stated above, the key point is the separation of the physics at 
the chosen energy scale from the physics at much higher energies; but how 
does this separation or ``decoupling'' exactly obtain?
In the QFT framework, the decoupling of physical phenomena, as well 
as the changing of the (effective) physical description as the 
scale changes, occur according to specific and precise rules. 
This is due essentially to some important peculiarities of the local 
quantum field description and, in particular, to the concept of the 
{\it renormalization group} (RG) and its deep impact on particle physics.
The emergence of the EFT idea (and approach) is in fact 
intertwined with the development of {\it renormalization theory} (RT): 
from the so-called old RT, introduced in the 1940s 
to deal with the divergence problems arising in quantum electrodynamics, 
to the recent new understanding of renormalization which is 
grounded on the RG concept. 
Without entering in detail, let us shortly see how EFT 
comes out in this context.\footnote{There is now a quite rich 
literature on RT, its history and its conceptual development.
Standard references are Cao and 
Schweber (1993), and Brown (1993).}

In its original meaning, renormalization was nothing other than a 
means of removing the infinities occurring 
in perturbative calculations in a QFT. To do this, 
the conventional strategy adopted in the ``old RT'' 
was more or less the following: 
first separate the divergent parts (high energy processes) from the finite parts 
(low energy processes) in some way, usually by introducing a 
{\it cutoff} $\Lambda$ (threshold energy for the validity of the 
theory); then absorb the  
divergences in some appropriate redefinition (``renormalization'') of  
the parameters (such as masses and coupling constants) of the theory;   
finally, to take into account the neglected high energy effects (in 
a cutoff-independent theory), remove the cutoff 
by letting  $\Lambda \rightarrow \infty$ --- with the consequently 
arising problems as regards 
the actual meaning of the cutoff (on this last point, see 
especially Cao and Schweber, 1993, pp. 52-55). 

A first step towards a new understanding of RT was 
the introduction of the RG concept in the QFT framework 
during the 1950s, on the grounds 
of {\it renormalization invariance}:\footnote{The seminal works are 
due to E.C.G. Stueckelberg and A. Petermann in 1953, M. Gell-Mann and 
F. Low in 1954, and N.N. Bogoliubov and D.V. Shirkov in 1955.} i.e., an 
arbitrariness in the choice of the parametrization of the theory to renormalize 
(and the consequent introduction of RG as the group of transformations 
relating the different parametrizations).
That RT was not just a technical device for removing infinities 
became a clear and 
accepted idea only after the revival and extension of the application 
of the RG methods by Kenneth Wilson in the early 1970s. 
By applying his previous results on RG in field theories on a lattice 
to the study of critical 
phenomena (generalizing the 1966 Kadanoff theory of scaling 
near the critical point for an Ising ferromagnet), Wilson in fact laid 
the basis for what is now the current conception of renormalization:  
namely, the conception that renormalization is essentially  
`an expression of the variation of the structure of physical interactions 
with changes in the scale of the phenomena being probed' (Gross, 1985, p. 153). 
The RG enters the picture by regulating (through the so-called 
RG equations) the way in which this variation occurs. 
Let us just focus on some fundamental steps.

A first conceptual point is the realistic interpretation of the cutoff, 
that is to `take seriously the idea of a physical cutoff at 
a very large energy scale $\Lambda$' (Polchinski, 1984, p. 269). 
A second point is `the idea of smoothly lowering the cutoff' (p. 270). 
On the RG approach, the effect of changing the scale or 
{\it rescaling} ($\Lambda_{o}\rightarrow\Lambda(s)=s\Lambda_{o}$) can in fact 
be absorbed in a change of the parameters, so that, for one parameter $g$, 
a trajectory $g=g(s)$ is defined as $\Lambda(s)$ varies. The RG equations 
so describe the flow of the parameters in a parameter space as one changes 
the scale. The important thing is that, typically, as one scales 
down to lower energies the solutions of the RG equations 
approach a finite dimensional sub-manifold in the space of possible 
Lagrangians: thus generally defining an {\it effective low energy theory}, 
which is formulated in terms of a finite number of parameters and is largely 
independent of the high energy starting situation (to be more precise: 
independent up to high energy effects that are suppressed by powers of
 $E/\Lambda$, where $E$ is the low energy at which the effective theory 
is appropriate).

Practically, in terms of a situation with particles having very 
different masses, what results is the following: 
if the energy $E$ at which we are working (the experimental energy) 
is much less than the mass $M$ of a particle 
(the ``heavy'' particle), we can act as if the particle was not there. 
The mass $M$ plays the 
role of the cutoff and the physics at the energy $E{\ll}M$ is 
describable by an EFT which is approximately renormalizable. 
The nonrenormalizable interactions are due to the heavy particle that 
has been ignored and their effects are small, being represented by terms 
that are suppressed by powers of $E/M$. When, 
on the contrary, the experimental energy $E$ approaches the cutoff $M$, the heavy 
particle cannot be ignored any more (the effects of the nonrenormalizable 
interactions are no longer small). A new theory (``new physics'') 
is required, which can be a 
renormalizable QFT, another EFT or something completely different. 

Taking this view to its ``extreme version'' (Georgi, 1989, p. 455), 
we arrive at the following interesting scenario: in a situation 
with different elementary particles $p_{i}$, each particle Compton 
wavelength $\lambda_{i}$ ($\lambda_{i}=1/M_{i}$, 
where $M_{i}$ is the mass of the particle $p_{i}$) 
can be taken to be associated with a boundary between two EFTs.
For distances larger than $\lambda_{i}$ (or 
for energies lower than $M_{i}$), the particle $p_{i}$ is omitted from 
the physical description, which is obtained in terms of 
an effective theory with 
cutoff $\Lambda=M_{i}$; for distances shorter than $\lambda_{i}$ 
(energies higher than $M_{i}$), the particle $p_{i}$ is included and 
the physics is described by another theory, which can be 
an EFT with a higher energy 
cutoff (in case there is another particle with a larger mass).
 
In principle, we thus obtain a {\it level structure} of effective theories, 
each level corresponding to a given energy scale. 
In addition, these levels are {\it related} in a precise way, through 
the so-called matching conditions regulating the connections 
between the parameters in the two theories 
on either side of the boundary. Namely, the parameters must be so 
related that the two theories provide the same physical description 
just below the boundary.

Given this layered structure of EFTs, we can distinguish between 
two ways of looking at it:

a) The {\it way down}:\footnote{The terminology --- ``going up'' 
towards lower energies (larger distances)/``going down'' towards 
higher energies (smaller distances) --- is chosen in order to 
show the connection with the {\it way down/way up} distinction 
in the hierarchical conception of natural sciences discussed in 
Section 2.}
from the EFT appropriate at the available energy at which 
we start to the theories successively 
appropriate as we go down to higher energies. 

b) The {\it way up}: from the finer-grained theories at higher energies 
to the coarser-grained theories at lower energies. 

Low energy theories are approximations to the corresponding high 
energy ones: the ``way up'' accordingly marks the direction in 
which theories {\it emerge} 
one from each other, if we accept the `general idea of one theory 
$T_{1}$ being emergent from another $T_{2}$ if, in a certain part of $T_{2}$'s 
domain of application, the results of $T_{2}$ are well approximated 
by those of $T_{1}$' (Butterfield and Isham, 2000, p. 57).
 
\section{Conclusion: What the EFT Approach Suggests and What 
It Does not Imply}

Summing up: 

\begin{enumerate}

\item[$\bullet$] Current QFTs, once the best candidates for being 
fundamental theories, are now commonly seen as EFTs, 
i.e. low energy approximations to other (more fundamental) theories.

\item[$\bullet$] EFT is the most appropriate and convenient 
way of describing (in the framework of QFT) the relevant physics in a limited 
energy domain. It is therefore an intrinsically approximate 
and context-dependent description.

\item[$\bullet$] The EFT approach is grounded on the RG concept: 
the variation of the (effective) physical description with the changing 
scale is described by the RG equations.

\item[$\bullet$] The EFT approach in its extreme version provides 
a level structure (``tower'') of EFTs, each theory connected with 
the preceding one (going ``up'' in the tower) by means of the RG equations 
and the matching conditions at the boundary; the 
boundary for an EFT is set at the cutoff given by the mass $M_{i}$
of the next heaviest 
particle $p_{i}$ excluded in the EFT's physical description.

\item[$\bullet$] The way up in the level structure of EFTs marks 
the direction in which the theories emerge one from each other.

\end{enumerate}

In the light of these points, let us now turn to the conclusive point 
of this paper: whether and 
how the EFT approach is of some philosophical 
relevance, especially with regard to reductionism and related issues. 
We shall distinguish between: a) a positive part --- what the EFT approach 
can suggest; and b) a negative part --- what 
the EFT approach does not imply.  
 
a) EFTs undoubtedly offer a concrete example in the context of basic science 
for analysing the way in which theories (on different levels) can be 
related. More precisely, the EFT approach provides a level structure of 
theories, where the way in which a theory emerges from another (in the 
sense mentioned at the end of Section 4) is 
in principle describable by using the RG methods and the matching conditions 
at the boundary. The basic question concerning the inter-level 
relationships --- given a level structure, 
how are the units in different levels related? --- can here be addressed 
in a concrete and definite manner: we have formal and substantial 
tools for determining how successive effective theories 
are related one to each other. Moreover, what is particularly advantageous 
from the viewpoint of the philosophical discussion, 
the conceptual framework remains always the same. All the theories 
are formulated in the same QFT language, thus allowing us to avoid 
the typical translation problems arising when discussing 
``heterogeneous'' intertheoretical relationships. 

The tower picture 
grounded on the extreme version of EFT might appear to be 
mere speculation. It results, however, from generalizing a 
concrete situation: in other words, it is not ``pure'' speculation.
Today's physics (particle 
physics, first of all, but also other fields such as condensed matter physics) 
does in fact provide a number of concrete examples of known theories which are 
effective low energy versions of other known theories --- 
for example, the chiral effective theory SU(2)$_{R}$xSU(2)$_{L}$ 
approximating quantum chromodynamics (QCD) at low energies, 
or the Landau theory of Fermi liquids, which is 
the effective field theory of the low-energy excitations in a 
conductor.\footnote{An introductive lecture on the effective low energy 
theory of QCD is, for 
example, Gasser (1997). For a specific presentation of 
the Landau theory of Fermi liquids as an EFT see Polchinski (1992).}

b) As mentioned in Section 1,  
a discussion has recently arisen among physicists, 
historians and philosophers of physics 
about what the EFT approach entails with regard to reductionism 
and fundamentality in natural science. 
In particular, on the basis of EFT and its extreme version, definite 
conclusions have been drawn about 
whether a fundamental (final) theory exists, whether the 
notion of fundamentality in science makes sense and, finally, 
whether reductionism (or antireductionism) holds. 
The best example in this sense is provided by Cao and Schweber, 
who surely go further than anybody else in analysing the implications 
of EFT and in deriving specific philosophical theses from it.     
We already mentioned the kind of 
antireductionist position they defend. Their position is more 
precisely specified in the following terms: the EFT approach 
endorses `a pluralism in theoretical 
ontology, an antifoundationalism in epistemology and an antireductionism 
in methodology' (Cao and Schweber, 1993, p. 69). Central facts to which 
Cao and Schweber make reference for their theses are: first, the 
remarkable stability of an effective low energy theory (largely decoupled 
from the corresponding high energy theory); second, the necessity for an 
empirical input in determining the effective low energy theory 
(more precisely: in determining the coupling constants in the 
cut-off Lagrangian) in the case where the high energy theory is not known.

A specific discussion of the theses defended by Cao and Schweber  
is not our aim here.\footnote{A short critical 
discussion is provided in Huggett and Weingard (1995, pp. 187-189).} 
We are interested rather in the following point: 
how far are claims of the above sort legitimate? Let us attempt an answer 
by considering, in the light of what we have seen in Section 4, 
what the present EFT approach in particle physics 
does not imply as regards the reductionism and 
fundamentality issues. Schematically:

\begin{enumerate}

\item[$\bullet$] The fact that current QFTs are now seen as 
EFTs does not imply any specific thesis about the 
existence of a final theory. The high energy 
theory could be another EFT, a local QFT or something completely different 
(in this sense the picture of a tower of EFTs is in some ways 
misleading).\footnote{See for example the 
argument by Cao (1997, p. 347), according to which the endless character of 
the tower of EFTs is entailed by the local operator formulation of 
QFT.}

\item[$\bullet$] The EFT approach does not imply that the idea of a theory 
being more fundamental than another is meaningless. 
The fact that an effective low energy theory `is equally fundamental 
to the original high energy formulation insofar as our only concern 
is low energy physics' (Gross, 1999, p. 59) does not entail that the 
high energy theory could not be seen as more fundamental from a different 
perspective.

\item[$\bullet$] The EFT approach does not imply antireductionism, 
if antireductionism is grounded on the fact of emergence, 
as in the case of Anderson (1972) or Mayr (1988). 
The EFT schema, by allowing definite connections between the theory levels, 
provides an argument against the basic antireductionist claim 
of the scientists' debate. A reconstruction (the way up) is not excluded, 
even though it may have to be only in principle.\footnote{Not having the complete 
renormalizable theory at infinitely short distances, 
we need empirical inputs. But if such a theory was available, 
`we could work our way up to the effective theory 
at any larger distance in a totally systematic way' (Georgi, 1989, p. 455). 
In general, as already said, if the high energy theory is not known, we need 
an empirical imput in determining the low energy effective theory.} 
In this sense, EFT does not represent 
a vindication of Anderson's 1972 views, as has been claimed 
(Schweber, 1993, p. 36).

\end{enumerate}

In short: keeping ideology away, EFT does not provide general answers to the 
basic questions of the ``scientists' debate''.    
The present EFT approach in particle physics 
is actually a practical and convenient way of proceeding 
in describing natural phenomena, grounded on a very natural idea  --- 
the idea that the physical description depends on the scale --- 
which in the context 
of the modern view of QFT proves to be very useful and fruitful. 
Moreover, the known EFTs surely offer new and interesting concrete cases 
for the philosophical discussion on inter-level relationships. \\

\noindent{\footnotesize{\it Acknowledgements}---The idea of writing this paper 
emerged from a discussion with the physicist Andrea Cappelli; I am 
grateful for his suggestions and initial encouragement. 

Earlier versions or portions of this paper were presented 
at the University of Oxford, the University of Florence, the SILFS'99 
Conference in Urbino, the University of Padoa, and 
the 11$^{th}$ International Congress of Logic, 
Methodology and Philosophy of Science in Cracow. I thank the audiences for 
stimulating questions. 

It is a pleasure to thank Jeremy Butterfield for his very 
helpful comments and advice. Thanks also to one anonymous referee for useful 
remarks.}\\\\

\begin{center}
{\bf References}
\end{center}

\noindent Anderson, P.W. (1972) `More Is Different', {\it Science} {\bf 177}, 
393-396.\\

\noindent Anderson, P.W. (1995) `Historical Overview of the Twentieth Century 
Physics', in L.M. Brown, A. Pais, and B. Pippard (eds.), {\it Twentieth Century 
Physics} (New York: American Institute of Physics Press), pp. 2017-2032.\\

\noindent Brown, L.M. (ed.) (1993) {\it Renormalization. From Lorentz to Landau 
(and Beyond)} (New York: Springer).\\

\noindent Butterfield, J. and Isham, C.J. (1999) `On the Emergence of Time in 
Quantum Gravity', in J. Butterfield (ed.), {\it The Arguments of Time} 
(Oxford: Oxford University Press), pp. 111-168.\\

\noindent Butterfield, J. and Isham, C.J. (2000) `Spacetime and the Philosophical 
Challenge of  Quantum Gravity', in C. Callender and N. Huggett (eds.), 
{\it Physics meets Philosophy at the Planck Scale} 
(Cambridge: Cambridge University Press), pp. .\\

\noindent Cao, T.Y. and Schweber, S.S. (1993) `The Conceptual Foundations 
and the Philosophical Aspects of Renormalization Theory', {\it Synthese} 
{\bf 97}, 33-108.\\

\noindent Cao, T.Y. (1997) {\it Conceptual Developments of 20th Century 
Field Theories} (Cambridge: Cambridge University Press).\\

\noindent Cao, T.Y. (1999) `Why are we philosophers interested in quantum 
field theory?', in T.Y. Cao (ed.), {\it Conceptual Foundations of 
Quantum Field Theory} (Cambridge: Cambridge University Press), pp. 28-33.\\

\noindent Cat, J. (1999) `The physicists' debates on unification in 
physics at the end of the 20th century', {\it Historical Studies in the 
Physical and Biological Sciences} {\bf 28}, 253-300.\\

\noindent Gasser, J. (1997) `QCD at Low Energies', in F. Cornet and 
M.J. Herrero, {\it Advanced School on Effective Theories} (Singapore: World 
Scientific), pp. 1-42.\\

\noindent Georgi, H.M. (1989) `Effective quantum field theories', in P. Davies 
(ed.), {\it The New Physics} (Cambridge: Cambridge University Press), pp. 446-457.\\

\noindent Georgi, H. (1997) `Topics in Effective Theories', in F. Cornet and 
M.J. Herrero, {\it Advanced School on Effective Theories} (Singapore: World 
Scientific), pp. 88-122.\\

\noindent Glashow, S.L. and Lederman, L.M. (1985) `The SSC: A machine for the 
nineties', {\it Physics Today}, March, 28-37.\\

\noindent Gross, D. (1985) `Beyond Quantum Field Theory', in J. Ambj{\o}rn, 
B.J. Durhuus, and J.L. Petersen (eds.), {\it Recent Developments in Quantum Field 
Theory} (Amsterdam: Elsevier), pp. 151-168.\\

\noindent Huggett, N. and Weingard, R. (1995) `The Renormalization Group and 
Effective Field Theories', {\it Synthese} {\bf 102}, 171-194.\\

\noindent Humphreys, P. (1997) `How Properties Emerge', {\it Philosophy 
of Science} {\bf 64}, 1-17.\\

\noindent Klee, R. L. (1984) `Micro-Determinism and Concepts of Emergence', 
{\it Philosophy of Science} {\bf 51}, 44-63.\\

\noindent Mayr, E. (1982) {\it The Growth of Biological Thought} (Cambridge, 
Mass.: Harvard University Press).\\

\noindent Mayr, E. (1988) `The limits of reductionism', {\it Nature} {\bf 331}, 
475.\\

\noindent Polchinski, J. (1984) `Renormalization and Effective Lagrangians', 
{\it Nuclear Physics} {\bf B231}, 269-295.\\

\noindent Polchinski, J. (1992) `Effective Field Theory and the Fermi Surface', 
hep-th/9210046.\\

\noindent Sarkar, S. (1992) `Models of Reduction and Categories of Reductionism', 
{\it Synthese} {\bf 91}, 167-194.\\

\noindent Sarkar, S. (1998) {\it Genetics and reductionism} (Cambridge: Cambridge 
University Press).\\

\noindent Schweber, S.S. (1993) `Physics, Community and the Crisis in 
Physical Theory', {\it Physics Today}, Nov., 34-40.\\ 

\noindent Schweber, S. (1997) `A Historical Perspective on the Rise of the 
Standard Model', in  L. Hoddeson, L. Brown, M. Riordan, and 
M. Dresden (eds.), {\it The Rise of 
the Standard Model} (Cambridge: Cambridge University Press), pp. 645-684.\\  

\noindent Weinberg, A. (1963) `Criteria for Scientific Choice', {\it Minerva} 
{\bf 1}, 159-171.\\

\noindent Weinberg, A. (1964) `Criteria for Scientific Choice II: The 
Two Cultures'', {\it Minerva} {\bf 3}, 3-14.\\

\noindent Weinberg, S. (1965) `Why Build Accelerators?', in L.C.L. Yuan (ed.), 
{\it Nature of Matter. Purposes of High Energy Physics} (Brookhaven National 
Laboratory), pp. 71-73.\\

\noindent Weinberg, S. (1987) `Newtonianism, reductionism and the art of 
congressional testimony', {\it Nature} {\bf 330}, 433-437.\\

\noindent Weinberg, S. (1993) {\it Dreams of a Final Theory} (London: Random House).\\

\noindent Weinberg, S. (1995a) `Reductionism redux', {\it New York Times 
Review of Books}, October, 39.\\ 

\noindent Weinberg, S. (1995b) {\it The Quantum Theory of Fields} (Cambridge: 
Cambridge University Press), Vol. I.\\

\noindent Weinberg, S. (1997) `Changing Attitudes and the Standard Model', 
in L. Hoddeson, L. Brown, M. Riordan, and M. Dresden (eds.), {\it The Rise of 
the Standard Model} (Cambridge: Cambridge University Press), pp. 36-44.\\

\noindent Weinberg, S. (1998) `A Response ot Professor W\'ojcicki', {\it Foundations 
of Science} {\bf 1}, 79-81.\\

\noindent Weinberg, S. (1999) `What is quantum field theory and what 
did we think it was?', in T.Y. Cao (ed.), {\it Conceptual Foundations of 
Quantum Field Theory} (Cambridge: Cambridge University Press), pp. 241-251.\\

\noindent Weisskopf, V.F. (1965) `In Defense of High Energy Physics', in 
L.C.L. Yuan (ed.), {\it Nature of Matter. Purposes of High Energy Physics} 
(Brookhaven National Laboratory), pp. 24-27.

\end{document}